\documentclass[english,aps,preprint,showpacs,preprintnumbers,amsmath,amssymb]{revtex4}
\usepackage{graphicx,amsfonts}
\usepackage{epsfig}
\usepackage{dcolumn}
\usepackage{bm}
\begin{document}

\title{SU(5)$\otimes$Z$_{13}$ Grand Unification Model}
\vspace{0.3cm}

\author{Alex G. Dias$^1$\footnote{email: alex.dias@ufabc.edu.br},\, Edison 
T. Franco$^2$\footnote{email: edisontf@ift.unesp.br},\, 
Vicente Pleitez$^2$\footnote{email: vicente@ift.unesp.br}}

\affiliation{\vspace{0.3cm}\\ $^1$Centro de Ci\^encias Naturais e Humanas, Universidade Federal do ABC,\\ 
Rua Santa Ad\'elia 166, 09210-170 - Santo Andr\'e, SP, Brazil 
\vspace{0.3cm} \\ 
$^2$Instituto de F\'\i sica Te\'orica, Universidade Estadual Paulista \\ Rua Pamplona 145, 
01405-900 - S\~ao Paulo, SP, Brazil }

\begin{abstract}
We propose an $SU(5)$ grand unified model with an invisible axion and the unification of 
the three coupling constants which is in agreement with the values, at $M_Z$, of $\alpha$, 
$\alpha_s$, and $\sin^2\theta_W$. A discrete, anomalous, $Z_{13}$ symmetry implies that the
Peccei-Quinn symmetry is an automatic symmetry of the classical Lagrangian protecting, at the same 
time, the invisible axion against possible semi-classical gravity effects. Although the 
unification scale is of the order of the Peccei-Quinn scale the proton is stabilized by the fact that
in this model  the standard model fields form the  $SU(5)$ multiplets completed by new 
exotic fields and, also, because it is protected by the $Z_{13}$ symmetry. 

\end{abstract}

\pacs{12.10.Dm, 12.10.Kt, 14.80.Mz } 

\maketitle

\section{Introduction}
\label{sec:intro}
The unification idea, mainly in $SU(5)$~\cite{gg74}, is still an
interesting alternative for the physics beyond the standard model~\cite{nath07}.
Unfortunately, the minimal non-supersymmetric $SU(5)$ model  has been 
ruled out by experimental data: i) the proton is more stable than the prediction of the 
minimal model~\cite{sk99}; ii) the value of weak mixing angle at the 
$Z$-peak $\sin^2\theta_W(M_Z)=0.23122(15) $ [or alternatively $\alpha_s(M_Z)$] does not agree 
with experimental data~\cite{pdg}. It means that the three coupling constants do not 
meet at a single point if only the standard model particles are taken into account;
iii) the electron and $d$-like quark masses are equal at the 
unification scale, and iv) last but not least, neutrinos are massless in the model. 
Moreover, the supersymmetric version, i.e., SUSY $SU(5)$, although it allows an unification
of the coupling constants, it has serious problems with the proton decay~\cite{perez} (however
see \cite{roy}) and probably also with the electroweak data~\cite{boer}. Thus, it appears 
natural to ask ourselves if there are other options besides SUSY $SU(5)$ that yield 
convergence of the couplings, the observed value of the weak mixing angle  and the other 
parameters at the $Z$-pole, an appropriately stable proton and, at the 
same time, realistic fermions masses including neutrino masses. Another problem, not 
necessarily related to the previous one, concerns the existence of axions~\cite{pdgaxion}. 
Recently, the interest in theories involving such particles has raised also due experiments 
devoted to the search of axion-like particles~\cite{axionexp}.  If the axion does exist it 
is important to know the realistic model in which the Peccei-Quinn (PQ) symmetry can be automatically 
implemented and how the axion parameters can be stabilized against possible 
semi-classical gravitational effects~\cite{gravity}. 

On the other hand, it was shown in Ref.~\cite{321run} that in the
context of the multi-Higgs extension of the standard model with an invisible axion proposed in
Ref.~\cite{axionsm} we have: \textit{i)} the unification of the three gauge coupling constants 
near the PQ scale; \textit{ii)} the model predicts the correct value of the weak mixing 
angle at the $Z$-peak; \textit{iii)} the axion and the nucleon are stabilized by the cyclic 
$Z_{13} \otimes Z_3$ discrete symmetries; finally, \textit{iv)} although neutrinos got an arbitrary
Dirac mass, through the effective $d=10$ operators
$\Lambda^{-1}_{PQ}\Lambda^{-5}L\Phi_\nu L\Phi_\nu\phi^5$, the
left-handed neutrinos get also a Majorana mass $\leq2$ eV and the
right-handed neutrinos acquire a large Majorana mass term via $d=7$
effective operator $\Lambda^{-3}_{PQ}\,\overline{{\nu^c_{aR}}}(M_R)_{ab}\nu_{bR}(\phi^*\phi)^2$, 
implementing in this way a see-saw mechanism at the PQ energy scale.

Here we will consider an $SU(5)$ grand unified theory which unify
the model of Ref.~\cite{axionsm}, in such a way that the partner of the standard model fields 
in $SU(5)$ multiplets are new heavy fields. This model allows an stable proton, 
unification of the three coupling constants, a natural PQ symmetry of the 
classical Lagrangian, and the axion being protected against semi-classical gravity effects. 

The outline of the paper is as follows. In Sec.~\ref{sec:su5} we give the representation 
content of the model and the $Z_{13}$ and PQ charge assignments of the several multiplets. 
Next, in Sec.~\ref{sec:evolution}, we consider the running equations for the three gauge 
coupling constants related to the low energy $SU(3)_C\otimes SU(2)_L\otimes U(1)_Y$ symmetry 
group. In Sec~\ref{sec:proton} we consider the proton stabilization and other phenomenological 
consequences concerning  the model; finally the last section is devoted to our conclusions. 

\section{Non-SUSY $SU(5)$ grand unified theory}
\label{sec:su5}

In Ref.~\cite{axionsm} the representation content of the standard model was augmented by 
adding scalar fields and three right-handed neutrinos, in such a way that a discrete 
$Z_{13}\otimes Z_3$ symmetry was implemented in the model there. Explicitly, the particle 
content of the model is the following:
$Q_L=(u,\,d)^T_L\sim({\bf3},{\bf2},1/3)$, $L_L=(\nu,\,l)^T_L\sim({\bf1},{\bf2},-1)$
denote quark and lepton doublets, respectively; $u_R\sim({\bf3},{\bf1}, 4/3)$,
$d_R\sim({\bf3},{\bf1},-2/3)$, $l_R\sim({\bf1},{\bf1},-2)$,
$\nu_R\sim({\bf1},{\bf1},0)$ are the right-handed
components. It was also assumed that each charged sector gain mass from a different scalar 
doublet: $H_u$, $H_d$, $H_l$ and $H_\nu$ which generate Dirac masses 
for $u$-like, $d$-like quarks, charged leptons and neutrinos, respectively (all of them of 
the form $({\bf1},{\bf2},+1)=(\varphi^+,\,\varphi^0)^T$). Some other scalar fields were 
also considered in order to permit the full  symmetry realization:  a neutral complex 
singlet $\phi\sim({\bf1},{\bf1},0)$, a singly charged 
singlet $h^+\sim({\bf1},{\bf1},+2)$ and a triplet
$\vec{{\cal T}}\sim({\bf1},{\bf3},+2)$.  
Next, we wonder what is  the simplest  group 
embedding  the above representation content. The answer is: $SU(5)$. To achieve this, along 
with a $Z_{13}$ symmetry, we have to add new fermions and scalar fields. 

In this vein, the representation content of our $SU(5)$ model is as follows.   
For each family, the fermion representation content under
$SU(5)\supset SU(3)_C\otimes SU(2)_L\otimes U(1)_Y$, there are two
${\bf5}^*$:
$(\Psi^c)_{dL}=(d^c_1,\,d^c_2,\,d^c_3,\,E^-,\,-N)^T_L$, and
$(\Psi^c)_{eL}=(D^c_1,\,D^c_2,\,D^c_3,\,e^-,\,-\nu_e)^T_L$ and two
${\bf10}$:
\begin{equation}
\Phi_{dL}=\frac{1}{\sqrt 2}\left(
\begin{array}{ccccc}
0 & u^c_3 & -u^c_2 & -u_1 & -d_1\\
-u^c_3 &0 & u^c_1 & -u_2 & -d_2\\
u^c_2 & -u^c_1 & 0 & -u_3 & -d_3 \\
u_1 & u_2 & u_3 & 0 & -E^+ \\
d_1 & d_2 & d_3 & E^+ & 0
\end{array}
\right)_L,
\label{decu1}
\end{equation}
and
\begin{equation}
\Phi_{eL}=\frac{1}{\sqrt2}\,\left(
\begin{array}{ccccc}
0 & U^c_3 & -U^c_2 & -U_1 & -D_1\\
-U^c_3 &0 & U^c_1 & -U_2 & -D_2\\
U^c_2 & -U^c_1 & 0 & -U_3 & -D_3 \\
U_1 & U_2 & U_3 & 0 & -e^+ \\
D_1 & D_2 & D_3 & e^+ & 0
\end{array}
\right)_L,
\label{decu2}
\end{equation}
where $E$ and $N$ are heavy charged and neutral leptons, respectively, and $U,D$ 
are heavy quarks having the same electric charge of the respective quarks $u,d$. 
Finally, in the fermion sector we have to add fermionic neutral singlets 
$(N^c)_L\equiv N^c_L$ and $(\nu^c)_L\equiv \nu^c_L$. We have used a notation in 
which the subindex $e(d)$ denotes the multiplet to which the known leptons ($d$-like 
quarks) belongs to; on the other hand, the $u$-like quarks always belongs to the decuplet
$\Phi_d$. Notice that since the
known quarks and leptons belong to different representations of $SU(5)$, we have to 
impose that both quarks and leptons, and not quarks and anti-leptons, have gauge 
interactions through the left-handed components~\cite{note1}.

The scalars of the model are the usual ${\bf24}$, here
denoted by $\phi_{24}$, with vacuum expectation value (VEV) 
$\langle\phi_{24}\rangle~=~v_{24}~\textrm{diag}(1,1,1,-\frac{3}{2},
-\frac{3}{2})$;  a complex singlet $\phi_0$ which is almost the axion (we note 
that by considering a complex ${\bf24}$ it is possible to implement the axion in 
this model~\cite{georgi} however this may introduce troubles with proton decay). 
In order to break the $SU(2)\otimes U(1)$ symmetry and 
generate the fermion's Dirac masses we use four 
Higgs multiplets: two ${\bf5}$, and two ${\bf45}^*$ to avoid the prediction 
$m_e(M_U)=m_d(M_U)$ (The using of ${\bf45}$ for avoiding this mass relation was 
done in Refs.~\cite{45}.) Finally, we add a ${\bf10}$ ($D_{10}$) and a ${\bf15}$ 
($T_{15}$) which contains, respectively, the singlet $h^+$ and the triplet 
$\mathcal{T}$ of Ref.~\cite{axionsm}. $T_{15}$ gives Majorana masses to the 
active neutrinos. 
We will denote the ${\bf5}$ as $H^5_a=(h^1_a,\,h^2_a,\,h^3_a,\,h^+_a,\,h^0_a,)$ with 
$a=e,d$; and their VEVs are $\langle H^{5\alpha}_a\rangle=(v_{a5}/\sqrt2)\delta^\alpha_5$; 
the ${\bf45}^*$ will be denoted by $H^{45}_a\equiv (H^{45}_a)^{\alpha\beta}_{\rho}$; 
($H^{45}_a)^{\alpha\beta}_\rho=-(H^{45}_a)^{\beta\alpha}_\rho;\, 
(H^{45}_a)^{\alpha\beta}_\alpha=0$, with $\langle H^{45}_a
\rangle^{\alpha\beta}_\rho~=~(v_{a45}/\sqrt2)(\delta^\alpha_\rho-4\delta^4_\rho
\delta^\alpha_4)\delta^\beta_5$; finally, $\langle T^{\alpha\beta}_{15}\rangle=(v_{15}/\sqrt2)
\delta^\alpha_5\delta^\beta_5$. The decuplet $D_{10}$ does not necessarily get a VEV at 
lowest order. Since in this model all scalar's VEVs are of the order of the electroweak 
scale, except $\phi_{24}$ and $\phi_0$ which have VEVs of the order of  grand unified theory 
(GUT) and PQ scale, respectively, we have still the
hierarchy problem. It is only ameliorated because the GUT scale
is lower (as we will show below) than  in other grand unification models.

Consider the following Yukawa interactions,
\begin{eqnarray}
-{\cal L}_Y&=& \overline{(\Psi_e)_R}\;[G_{e5}\,\Phi_{eL}H^{5*}_e
+G_{e45}\,\Phi_{eL}H^{45}_e+G_{\nu} \nu^c_LH^5_e]+\overline{(\Phi^c_e)_R}\,
\epsilon\, K_U\,\Phi_{eL}H^{45*}_e
\nonumber \\
&+& \overline{(\Psi_d)_R}\;[ G_{d5}\,\Phi_{dL}H^{5*}_d+
G_{d45}\,\Phi_{dL}H^{45}_d +G_{N} N^c_L H^5_d]+\overline{(\Phi^c_d)_R}\,
\epsilon\, K_d\,\Phi_{dL}H^{45*}_d
\nonumber \\
&-&  \overline{(\Phi^c_e)_R}\,\epsilon\,F_U\,\Phi_{eL}H^5_e
-\overline{(\Phi^c)_{dR}}\,\epsilon\,F_d\,\Phi_{dL}H^{5}_d
- 
\overline{(\Psi_e)_R}\,G_{e15} \Psi^c_{eL}T_{15}
+H.c.,
\label{yukawa}
\end{eqnarray}
\noindent
where $G,K,F$ are $3\times3$ complex matrices but we have omitted generation 
and $SU(5)$ indices; $\epsilon$ denotes the 
$SU(5)$ fully antisymmetric tensor. With Eq.~(\ref{yukawa}) we obtain the mass matrices
($T$ denotes the transpose matrix)
\begin{eqnarray}
M_e=G^T_{e5} \,\frac{v^*_{e5}}{2}-3G^T_{e45}\,v_{e45},\quad
M_D=G_{e5}\frac{v^*_{e5}}{2}+G_{e45}v_{e45},\nonumber 
\\ M_U=\sqrt{2}v_{e5}(F_U+F^T_U)+2\sqrt{2}v^*_{e45}(K^T_U-K_U),
\label{mlmD}
\end{eqnarray}
and
\begin{eqnarray}
M_E=G^T_{d5}\frac{v^*_{d5}}{2}-3G^T_{d45}v_{d45},\quad
M_d=G_{d5}\frac{v^*_{d5}}{2}+G_{d45}v_{d45},\nonumber \\
M_u=\sqrt{2}v_{d5}(F_d+F^T_d)+
2\sqrt{2}v^*_{d45}(K^T_d-K_d),
\label{mEdu}
\end{eqnarray}
$M^{Dirac}_\nu=( v_{e5}/\sqrt2)G^T_\nu$, and
$M^{Dirac}_N=~(v_{d5}/\sqrt2)G^T_N$. The left-handed neutrinos have
a Majorana mass term coming from the $T_{15}$: $M^{Majorana}_\nu=(v_{15}/\sqrt2)G^T_{e15}$.
Both $v_{a5}$ and $v_{a45}$ are of the order of the electroweak scale, in fact
$\sum_a(\vert v_{a5}\vert^2 +\vert v_{a45}\vert^2)+\vert v_{15}\vert^2=(246\,\textrm{GeV})^2,
\;a=e,d$, with $\vert v_{15}\vert< 3.89$~GeV~\cite{comment}. For instance, using only one 
generation, if $v_{e5}=v_{e45}\equiv v_e$, assuming that these VEV are real and neglecting 
$v_{15}$, we have, from Eqs.~(\ref{mlmD}) and (\ref{mEdu}), $M_e=(G^T_{e5}/2-3G^T_{e45})v_e$
and $M_D=(G_{e5}/2+G_{e45})v_e$ (and similarly for $M_E$ and
$M_d$), so we can choose the Yukawa coupling constants must be such
that $M_e\ll M_D$, $M_u\ll M_U$ and $M_d\ll M_E$. In the context of three generations
all these mass matrices are $3\times3$ matrices and those relations among the masses 
refer to the respective eigenvalues. Right-handed components of neutrinos and the neutral 
leptons $N_R$, get also a Majorana mass term through the interactions with the 
axion~\cite{nusaxion}. 

We see that the representation content of the model implies that the 
vector bosons do not induce, at 
the tree level, the nucleon decay because these interactions involve the usual 
quarks and heavy leptons; or heavy quarks with the usual leptons. The same is true 
for the Yukawa interactions if they are given only by these in Eq.~(\ref{yukawa}). 
This diminish the importance of the constraints coming from nucleon decay on the 
leptoquarks masses. Thus, they may have a mass lower than the unification scale. 
Notwithstanding, when studying the evolution of the coupling constants, we
will assume that all leptoquarks are heavy enough and do not consider them 
in the running of the couplings. Next, we will show that the Yukawa interactions in 
Eq.~(\ref{yukawa}) are the only ones allowed by an appropriate discrete symmetry.

Let us use the fact that a $Z_N$ symmetry with $N$ being a 
prime number does not have any subgroup, in other words, it cannot be decomposed 
as $Z_p\otimes Z_q,\;(p,q<N)$, so that the  $Z_N$ symmetry may be a subgroup 
of a unique local group $U(1)$. In this vain, let us introduce the following $Z_{13}$ 
symmetry in the Yukawa interactions in such a way that only these interactions
in Eq.~(\ref{yukawa}) are allowed. The fields of the model transform under 
$Z_{13}$ as follows:
\begin{eqnarray}
(\Psi^c)_{eL}\to \omega_3 (\Psi^c)_{eL},\;(\Psi^c)_{dL} \to
\omega^{-1}_1(\Psi^c)_{dL},\;
\Phi_{eL}\to \omega^{-1}_1 \Phi_{eL},\;\Phi_{dL}\to
\omega^{-1}_4\Phi_{dL}, \nonumber \\
\nu^c_L\to \omega^{-1}_5\nu^c_L,\; N^c_L\to \omega_6
N^c_L,
\;H^{5}_e \to \omega_2H^{5}_e,\;
H^{45}_e \to \omega^{-1}_2 H^{45}_e, \;
H^{5}_d \to \omega^{-1}_5 H^{5}_d,\nonumber \\
H^{45}_d \to \omega_5 H^{45}_d,\;
D_{10}\to\omega_3 D_{10},\; T_{15}\to \omega^{-1}_6
T_{15},\; \phi_{24}\to \omega_0\phi_{24},\; \phi_0\to
\omega_4\phi_0. \label{z13}
\end{eqnarray}
We have assumed that the three generations are replicas under $Z_{13}$. However, it 
could be interesting to consider the case when this is not the case.  

The scalar potential has hermitian quadratic terms $\mu^2_\chi\chi^\dagger\chi$ 
(where $\chi$ denotes any of the Higgs scalar multiplets of the model), 
which are needed to break the electroweak symmetry, trilinear and quartic Hermitian terms, and 
non-Hermitian self-interactions which are trilinears 
\begin{eqnarray}
H_e^5H_e^{45}\phi_{24},H^5_dH^{45}_d\phi_{24},(H^5_d)^2D^*_{10},
H^5_dH^{45*}_dD^*_{10},(H^{45}_d)^2D_{10}, \label{tri}
\end{eqnarray}
and quartic:
\begin{eqnarray}
&&H^5_dH^{45}_d\vert T_{15}\vert^2,H^5_dH^{45}_d\phi^2_{24},
H^5_dH^{45*}_dD^*_{10}\phi_{24},H^5_dH^{45}_dH^{5*}_e H^{45*}_e,
H^5_dH^{45*}_d(H^{5*}_d)^2,\nonumber
\\ &&
(H^{45}_e)^3T^*_{15},T_{15}D^*_{10}\phi_{24}\phi_0^*,
(H^{45}_d)^2D^*_{10}\phi_{24},(H^5_e)^2H^{45*}_eT_{15},
(H^{45}_d)^2T_{15}\phi_0^*,(H^5_e)^2H^{5*}_eH^{45}_e, \nonumber \\
&&(H^5_e)^2(H^{45}_e)^2,
(H^5_d)^2(H^{45}_d)^2,H^5_eH^{45}_e\vert H^{5}_d\vert^2,
 H^5_eH^{45}_e\vert H^{45}_e\vert^2,
H^5_eH^{45}_e\vert H^{45}_d\vert^2,\nonumber
\\ && H^5_eH^{45}_e\vert D_{10}\vert^2,
H^5_eH^{45}_e\vert T_{15}\vert^2,H^5_eH^{45}_e\phi^2_{24},
\vert H^5_e\vert^2H^5_dH^{45}_d,
H^5_e(H^{45*}_e)^2T_{15},H^5_eH^{45}_eH^5_dH^{45}_d,\nonumber \\ && (H^5_d)^2H^{5*}_dH^{45}_d,
(H^5_d)^2D^*_{10}\phi_{24},(H^5_d)^2T^*_{15}\phi_0,
H^5_dH^{45}_d\vert H^{45}_e\vert^2,H^5_dH^{45}_d\vert
H^{45}_d\vert^2, H^5_dH^{45}_d\vert D_{10}\vert^2.\label{4ticos}
\end{eqnarray}

Moreover, with the interactions in Eq.~(\ref{yukawa}) and the non-hermitian interactions 
in (\ref{tri}) and (\ref{4ticos}), allowed by the symmetry in Eq.~(\ref{z13}), the 
PQ symmetry is automatic. The PQ charges are shown between parenthesis in units of the PQ charge of $\Psi_d$ as follows:  
\begin{eqnarray}
& & (\Psi^c)_{eL}(-1/3),\,\, (\Psi^c)_{dL}(1), \,\, \Phi_{eL}(1/9), \,\, \Phi_{dL}(-1/3), 
\,\,\nu^c_L(5/9), 
\nonumber \\
& & N^c_L(-5/3), \,\, H^5_e(-2/9),\,\,   H^{45}_e(2/9),\,\, H^5_d(2/3),\,\, H^{45}_d(-2/3), 
\nonumber \\
& & D_{10}(4/3), \,\, T_{15}(2/3), \,\, \phi_{24}(0), \,\, \phi_0(-2/3). 
\label{pqc2}
\end{eqnarray}

As in the model of Ref.~\cite{axionsm}, the $Z_{13}$ protect the axion against possible 
semi-classical gravity effects. The model has no domain wall problem~\cite{kim}.

\section{Evolution of the coupling constants}
\label{sec:evolution}

In order to study the running of the coupling constants in a consistent way with the present
model, we augmented the representation content of the model of Ref.~\cite{axionsm}. Hence, we 
assume that the only extra degrees of freedom that are active at low energies, i.e., below the
unification scale but above the electroweak scale, transforming under the standard model (SM)
symmetries are (per family): $(N,\,E)^T_L\sim({\bf3},{\bf2},-1)$,
$(U,\,D)^T_L\sim({\bf3},{\bf2},1/3)$ and the respective singlets 
$E_R\sim(\textbf{1},\textbf{2},-2)$, $N_R\sim(\textbf{1},\textbf{1},0)$, $U_R\sim({\bf3},
{\bf2},2/3)$ and $D_R\sim({\bf3},{\bf1},-1/3)$. In the Higgs boson sector we add also four 
scalar doublets, two of them $H_d$, $H_l$ are those that belong to ${\bf5}$ and two others, say 
$H_u$ and $H_\nu$ which belong to the ${\bf45}$; a triplet $\mathcal{T}$ belonging
to $T_{15}$ and the singlet $h^+$ which is part of the $D_{10}$. As in Ref.~\cite{321run} only
$h^+$ will be considered with mass of the order of the unification scale. 

Let us look at the evolution equations at the 1-loop approximation, with 
all the new fermions entering only above an intermediate energy scale $ \mu_{_{IS}}$ which is
certainly bigger than the electroweak scale. Some of the new fermions could have mass below the
known heavier Standard Model particles,  but we will not consider such possibility. 
Below we comment more on that (see Sec.~\ref{sec:con}). Thus, the 1-loop equations are: 
\begin{equation}
\frac{1}{ \alpha_i(\mu)}= \frac{1}{\alpha_i(M_{_Z})}-\frac{1}{2\pi}  
\left[ b_i\ln \frac{\mu_{_{IS}}}{M_{_Z}} + b^{^{IS}}_i\ln \frac{\mu}{\mu_{_{IS}}} \right], 
\label{rg1loop}
\end{equation}
where $\alpha_i(M_{_Z})=g_i^2(M_{_Z})/4\pi$ are the usual gauge couplings defined for these 
equations, and $b_i$ the well know coefficients for a general 
$SU(N)$ gauge group,  given by $b_i~=~(2/3)\sum T_{Ri}(F)+(1/3) \sum T_{Ri}(S)-(11/3)\,
C_{2i}(G)$ for Weyl fermions ($F$) and complex scalars ($S$), and $T_R(I)\delta^{ab}=
\textrm{Tr}\{T^a(I),T^b(I)\}$ with $I=F, S$; $T_R(I)=1/2$ for the fundamental representation, 
$C_2(G)=N$ when $N\geq 2$, for $U(1)$, $C_2(V)=0$, and $T_{R1}(S_a,F_a)=(3/5)\textrm{Tr}(Y^2_a/4)$.  
The same is valid for the $b^{^{IS}}_i$ with the counting extending to the exotic 
fermions representations.  At the $SU(3)_C\otimes SU(2)_L\otimes U(1)_Y$ energy level with 
$N_g$ fermion generations, $N_H$ scalar doublets ($Y=\pm1$) and $N_T$ non-hermitian scalar 
triplets ($Y=2$), and $N_s$ charged singlets, we have:
\begin{equation}
b_1=\frac{4}{3}N_g+\frac{1}{10}N_H+\frac{3}{5}N_T+\frac{1}{5}N_s,\quad b_2= 
\frac{4}{3}N_g+\frac{1}{6}N_H+\frac{2}{3}N_T-\frac{22}{3}, \quad
b_3=\frac{4}{3}N_g-11,
\label{bi}
\end{equation}
where a grand unification normalization factor (3/5) for the hypercharge $Y$ assignment 
is included in $b_1$.
So that according the additional representations in the beginning of
this section only the heavy fermions are activated above $\mu_{_{IS}}$, we have (with $N_s=0$)
\begin{equation}
(b_1,b_2,b_3)=(5,-2,-7),\quad (b^{^{IS}}_1, b^{^{IS}}_2,b^{^{IS}}_3)=(9,2,-3). 
\label{bi2}
\end{equation}
Note that there is no asymptotic freedom in $\alpha_2$ at the 1-loop level. 
The grand unification mass scale and the weak mixing angle are given by
\begin{equation}
M_{GUT}=\mu _{_{IS}}\exp \left\{ 2\pi \frac{\left[ \frac{3}{5}\alpha
^{-1}\left( M_{Z}\right) -\frac{8}{5}\alpha _{3}^{-1}\left( M_{Z}\right) %
\right] }{\left( b_{1}^{^{IS}}+\frac{3}{5}b_{2}^{^{IS}}-\frac{8}{5}%
b_{3}^{^{IS}}\right) }\right\} \times \left( \frac{M_{Z}}{\mu _{_{IS}}}\right) ^{%
\frac{\left( b_{1}+\frac{3}{5}b_{2}-\frac{8}{5}b_{3}\right) }{\left(
b_{1}^{^{IS}}+\frac{3}{5}b_{2}^{^{IS}}-\frac{8}{5}b_{3}^{^{IS}}\right) }},
\label{gut1}
\end{equation}
and 
\begin{equation}
\sin ^{2}\theta _{W}=\frac{3}{8}+\frac{5}{8}\frac{\alpha \left( M_{Z}\right)}
{2\pi }\left[ \left( b_{1}-b_{2}\right) \ln \left( \frac{M_{Z}}{\mu _{_{IS}}}%
\right) +\left( b_{1}^{^{IS}}-b_{2}^{^{IS}}\right) \ln \left( \frac{\mu _{_{IS}}}{%
M_{GUT}}\right) \right].
\label{sin1}
\end{equation}

However, since 
\begin{equation}
b_1-b_2=b^{^{IS}}_1-b^{^{IS}}_2,\;\;  b_{1}+\frac{3}{5}b_{2}-\frac{8}{5}b_{3}=
b_{1}^{^{IS}}+\frac{3}{5}b_{2}^{^{IS}}-\frac{8}{5}
b_{3}^{^{IS}},
\label{upa}
\end{equation}
$M_{GUT}$, and $\sin^2{\theta}_W(M_Z)$
defined at the  $Z$ boson mass, do not depend on the scale $\mu_{_{IS}}$ and we are left with
\begin{equation}
M_{GUT} =M_{Z}\exp \left[2\pi \frac{\alpha ^{-1}\left( M_{Z}\right) -
\frac{8}{3}\alpha _{3}^{-1}\left( M_{Z}\right) }{\frac{5}{3}b_{1}+b_{2}-
\frac{8}{3}b_{3}}\right],
\label{gut2}
\end{equation}
and
\begin{equation}
\sin ^{2}\theta _{W}(M_Z) =\frac{3}{8}+\frac{5}{8}\frac{\alpha \left(
M_{Z}\right) }{2\pi }\left( b_{1}-b_{2}\right) \ln \left( \frac{M_{Z}}{
M_{GUT}}\right).
\label{sin2}
\end{equation}
Both $M_{GUT}$ and $\sin^2\theta_W(M_Z)$, at the one loop level,
are the same as in Ref.~\cite{321run}. Using $M_Z~=~91.1876$ GeV; $\alpha(M_Z)=1/128$; and 
$\alpha_3(M_Z)=0.1176$~\cite{pdg},
we obtain $M_{GUT}~=~2.86~\times~10^{13}$ GeV and 
$\sin ^{2}\theta _{W}(M_Z)=0.23100$, in agreement with the 
usual value~\cite{pdg}. Moreover, using the evolution equations in Eq.~(\ref{rg1loop}) 
we get $\alpha^{-1}_{GUT}\simeq 23 (21)$, if $\mu_{_{IS}}\approx1$ TeV ($\mu_{_{IS}}=M_Z$).
The inclusion of the scalar singlet at low energies ($N_s=1$) gives worse values for 
this mixing angle so, it must be considered with mass near the unification scale.
In Fig.~1 we show the evolution of the coupling constants at the 1-loop level in the 
present model. An analysis at the 2-loop can be done, but in general  it does not lead to 
a prediction unless the top quark and all extra fermion and scalar fields are 
taken into account. This result in a large set of coupled equations~\cite{jones} that 
deserve more careful study.

\begin{figure}[ptb]
\begin{centering}
\includegraphics[width=8cm,height=7.0cm]{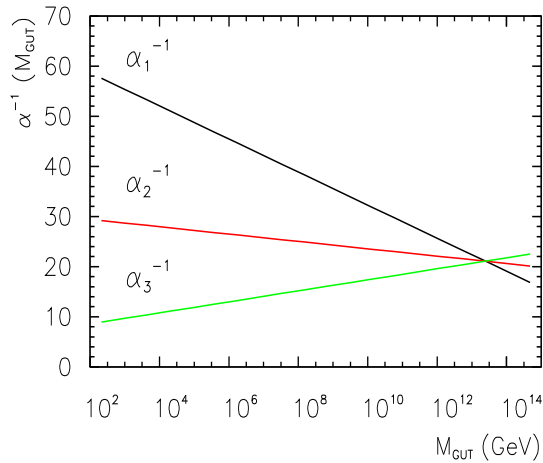}
\par
\end{centering}
\caption{ \label{fig1} In this figure we show the convergence
point for the SU(5) model here.}
\end{figure}

Notice that in this extension of the SM, the Yukawa interactions 
can be similar to those in Ref.~\cite{axionsm} but, it is worth noting that, if we want 
to avoid a general mixing in each charge sector the extra quark (lepton) generation must 
transform under $Z_{13}$ in a different way from those of the usual lepton (quarks). 
However, we recall that getting a small mixing it can be interesting if in the  future
a departure from unitarity in the Cabibbo-Kobayashi-Maskawa mixing matrix would be observed~\cite{kylee}. In 
fact, even the usual three generations can be transformed under $Z_{13}$ different from 
each other in such a way that predictive mass matrices can be obtained.

\section{Stabilizing the proton}
\label{sec:proton}

As we said before, in this model nucleon decays are forbidden at the tree level. Here we will
discuss this point in more detail. The effective operators with dimension six~\cite{weinberg}, 
$d=6$,  that can induce the proton disintegration 
do not operate in our model because vector leptoquarks always mix the usual fermions with the
heavy ones. However, without the $Z_{13}$ symmetry, there are still dangerous d=4 operators coming 
from the Yukawa couplings with the $H^{45}$  Higgs scalars. For instance, without that 
discrete symmetry, Yukawa interactions like $\bar{\Psi}_{eR}\Phi_{dL}H^{45}_e$ and 
$\bar{\Phi}^c_{dR}\Phi_{dL}\epsilon H^{45*}_e$ are allowed. These terms induce the proton decay
through interactions like $\bar{Q}^c_{mR}\epsilon \vec{\sigma}\cdot\vec{\eta}_m L_L$, and 
$\epsilon_{mnp}\bar{Q}^c_{mR}\epsilon \vec{\sigma} \cdot\vec{\eta}_nQ_{pL}$, respectively; 
here $\epsilon=i\sigma_2$, and $m,n,p$, are color indices, 
and $\vec{\eta}_m$ is the colored scalar triplet belonging to $H^{45}_e$. 
Once the $Z_{13}$ symmetry is introduced the Yukawa interactions allowed are just those 
given in Eq.~(\ref{yukawa}), and they only induce interactions like 
$\bar{Q}^c_{mR}\epsilon~\vec{\sigma}~\cdot~\vec{\eta}_m L^\prime_L$ and $\epsilon_{mnp}
\bar{Q}^c_{mR}\epsilon\vec{\sigma}\cdot\vec{\eta}_nQ^{\,\prime}_{pL}$, where the primed 
fields are heavy quarks, $U,D$, or heavy leptons, $E,N$. Hence, with the interactions in 
Eq.~(\ref{yukawa}), independently of the mixing in the scalar sector, the nucleon is not 
allowed to decay at the tree level. The model is in this respect phenomenologically safe. 

The $Z_{13}$ symmetry introduced in this model allows 
effective interactions with flavor changing neutral current. For instance, 
\begin{equation}
\frac{g^2_{Z_N} }{M^2_{Z_N}}\,
h_{abcd}\,\overline{L_{aL}}\,\gamma^\mu L_{bL} \overline{L_{cL}}\,\gamma_\mu L_{dL},
\label{a1}
\end{equation}
here $a,b,c,d$ are family indices, and $g_{Z_N}$ and $M_{Z_N}$ denote the coupling constant
of the $Z_N\subset U(1)_{local}$ and the mass of the (heavy) vector boson 
associated with this symmetry, respectively and $h_{abcd}$ are dimensionless constants 
(there are also effective interactions induced by the heavy scalar that condensate at very 
high energies). The interactions in Eq.~(\ref{a1}) induce rare transitions like $\mu\to eee$.
Neglecting the electron masses we can write the width of this decay in terms of the muon 
decay width as follows:
\begin{equation}
\Gamma_{\mu\to 3e}=\left( \frac{g_{Z_N}}{g_2}\,\frac{M_W}{M_{Z_N}}
\right)^4\, \Gamma^{\textrm{SM}}_{\mu \to e\nu\bar{\nu}},
\label{a2}
\end{equation}
where $g_2$ and $M_W$ are the well-known parameters of the standard model and we see that 
even if $g_{Z_N}\sim O(g)$ with $M_{Z_N}>10^3 M_W$, we have already got a suppression factor of
$10^{-12}$. However, it is more natural that $M_{Z_N}$ be of the order of the breakdown of 
the local $U(1)$ symmetry i.e., at least of the order of the PQ scale. It may be also 
interesting to assume that $g_{Z_N}\ll g$ at low energies, in such a way that, since 
$g_{Z_N}$ which is not an asymptotic free parameter it can fit with $g$ at a high energy 
and the other coupling constant of the low energy model. 

Just as another example, there are also interactions like
\begin{equation}
\frac{g^2_{Z_N}}{M^2_{Z_N}}\,
h^\prime_{abcd}\,\overline{Q_{aL}}
\gamma^\mu Q_{bL} \overline{F_{cL}}\gamma_\mu F_{dL},
\label{a3}
\end{equation}
where $F=Q,L$ and $h^\prime$ is another dimensionless matrix. When $F=Q$
this interaction will induce a contribution to $\Delta M_K$ and
other related parameters. Notice, however, that
\begin{equation}
\Delta M_K\propto \left( \frac{g_{Z_N}}{g}\,
\frac{M_W}{M_{Z_N}}\right)^2\,G_F B_Kf^2_Km_K,
\label{a4}
\end{equation}
we see that this contribution to $\Delta M_K$ is rather small for the same
values of the $Z_N$ parameters in Eq.~(\ref{a2}).

In general discrete symmetries may be not free of anomalies. 
Although it is interesting to looking for cyclic local discrete symmetries that are 
anomaly free, we would like to emphasize that it is not necessarily a loophole of models 
with anomalous $Z_N$ symmetries.
If $g_{Z_N}\leq g$ the transition violating $B$ and $L$ conservation
induced by the anomaly of the $Z_N$ symmetry will be smaller
than $e^{-16\xi \pi^2/g^2}\approx 10^{-117\xi}$~\cite{thooft}, with $\xi=O(1)$ a
model dependent parameter. Although this transition
is negligible at zero temperature it may be important in
earlier ages of the universe as a mechanism for leptogenesis generation through the 
decays of the heavy neutral leptons if CP violation is implemented in it. In fact, the model
allows several ways to implement baryogenesis and leptogenesis~\cite{baryo,lepto} as we will 
shown elsewhere.

\section{Conclusions}
\label{sec:con}

Summarizing, we have obtained an $SU(5)$ extension of a previous model of 
Refs.~\cite{axionsm},  which is  as good as SUSY $SU(5)$ with respect 
to the unification of the electroweak and strong interactions. We also have in this
model that the proton is stable and the PQ an automatic symmetry of classical 
Lagrangian,  with the invisible axion protected against gravitational effects by a local
$Z_{13}$ symmetry. The unification of the three coupling constants occurs at the PQ scale 
as in \cite{321run}. It is important to realize that the model does not admit supersymmetry 
at least at low energies, but we have seen that, in order to have unification, supersymmetry 
is not an indispensable factor anymore. As it was put forward in Ref.~\cite{nusaxion}, 
the PQ energy scale can be related with the mass of the sterile neutrinos~\cite{nusaxion}, 
and in the present model the PQ energy scale is related with the GUT scale.
Put all this together and we have that in the present model it is possible to have 
$M_{\nu_R}\ll M_{PQ}\sim M_{GUT}$ or $M_{\nu_R}\sim M_{PQ}\sim M_{GUT}$, depending on the 
$Z_{13}$ charges assignment.

Although the scalar and vector leptoquarks do not induce the nucleon decay at the tree level,
their masses are of the order of the unification scale (they gain masses from the 
$\langle\textbf{24}\rangle$). Since this scale is of the order of 
the PQ scale it means that both energy scale can be related to each other. 
On the other hand, the exotic quarks and leptons $U,D,E$ gain mass from $\textbf{5}$
and $\textbf{45}$ which have VEVs of the order of the electroweak energy scale and, for this reason, 
they could not be very heavy. We recall that the experimental limits on the exotic leptons and quarks 
like $E,N$ and $U,D$, respectively, are model dependent but since all of them gain mass from VEVs 
of the order of the electroweak scale they
must not be very heavy indeed. For instance, from data we have lower
bounds on the masses (in GeV) of a possible fourth family~\cite{pdg}: 
for sequential $E^{\pm }$ charged lepton, we have $m>100.8$, CL=$95\%$  
(decay to $\nu W$); for stable charged heavy leptons $m>102.6$,~C.L.=$95\%$; for 
stable neutral heavy lepton the limits are
$m>45.0$, C.L.=$95\%$ (Dirac) and $m>39.5$, C.L.=$95\%$ (Majorana). Finally, for extra 
quarks of the $b$-type ($b^{\prime }$ 4th generation) the lower limits
are $m>190$, C.L.=$95\%$ 
(quasi-stable $b^{\prime }$) or $m>199$, C.L.=$95\%$ (neutral currents); if it decays 
in $ll$ +jets, $l$+jets, we have
$m>128$, C.L.=$95\%$. Of course, these limits are strongly model dependent (in some 
models the fourth family is almost degenerate~\cite{datta}). On the other hand, as 
we said before, the model has a general 
mixing among the fields of the same electric charge sector, thus the 
generalized unitarity triangle analysis of the Cabibbo-Kobayashi-Maskawa 
matrix~\cite{kylee} can be used for deriving upper bounds on the coefficients of 
the effective operators inducing such mixings~\cite{ckm}.

The model has right-handed neutrinos, so it is possible that an
$SO(10)$ would be more appropriate for the unification of the model of 
Ref.~\cite{axionsm}. However since $SU(5)\subset SO(10)$ our
$SU(5)$ model is already good enough for implementing a GUT theory for the 
extension of the standard model with proton stability. 
 
\acknowledgments E.T.F. was supported by FAPESP and V. P. was
supported partially by CNPq under the processes  03/13869-3 and 300613/2005-9, 
respectively. A.G.D. also thanks FAPESP for financial support.


\begin{thebibliography}{99}
\bibitem{note1} In the standard model and its extensions we do not know why the up and down quark 
have to go together with electron and its neutrino. This arrangement have no experimental basis 
and it just reflects the order in which we met the particles~\cite{quigg}.
\end{thebibliography}
\end{document}